# Instant messaging an effective way of communication in workplace

Tirus Muya Maina,
Murang'a University College
ICT Directorate
tmuya@mruc.ac.ke

## Abstract

The modern workplace is inherently collaborative, and this collaboration relies on effective communication among co-workers. Instant messaging is the multi-tasking tools of choice most people chatting over IM do other things at the same time. The use of IM in workplace is less intrusive than the use of phone, more immediate than email and has added advantage due to the ability to detect presence. In order for institution to maximize increased business productivity using instant messaging it's imperative that organizations define and publish ICT policies, guidelines and regulations.

Overall IM boosts business performance by making operations faster, more agile, and more efficient with very little additional cost thus Organizations that deploy IM would reap significant Return on Investment. Institutions should adopt IM meetings which are be more efficient and less prone to straying off topic, because of the relative effort of typing versus talking.

**Keyword:** Instant Messaging, IM, ICT, Communication

## Introduction

Informal face-to-face (FTF) communication has been shown to serve many important functions in organizations, including complex coordination, problem solving, and social learning (Whittaker, Frohlich, & Daly-Jones, 1994). Early attempts to build tools to support informal communication focused on audio and video environments. However, these attempts have not been widely adopted for several reasons, including the lack of support for core user tasks, cost, privacy concerns, and implementation difficulties (Whittaker, 1995). Instant Messaging (IM), in contrast, has become of great interest to the modern community because it is a tool that successfully supports informal communication (Nardi, Whittaker, & Bradner, 2000).

The modern workplace is inherently collaborative, and this collaboration relies on effective communication among co-workers. Many communication tools; email, blogs, wikis, Twitter, etc. have become increasingly available and accepted in workplace communications (Turner, Qvarfordt, Biehl, Golovchinsky, & Back, 2010). According to Woodard (2011) she noted that with the introduction of the Internet, much of the communication has been done through the screen of our computer monitors or even through our phones. She further notes that the world has become a small global village, and the Internet is the superhighway that connects every house, organisation and institution in this fist-sized world today.

D'Silva (2013) observes that "Apart from being a portal for information, entertainment and education, the Internet plays another very integral role that of a communication device". The Internet has today been proved to be one of the most economic way that people across continents can communicate with each other. Email and instant messaging have today taken the place of courier and fax as reliable corporate communication devices. One of the many additions to this was that of instant messaging. Instant messaging, or IMs, allows for instant communication to anyone in the same city, in a different state, or anywhere in the world (D'Silva, 2013).

Reported cases where IM was preferred to informal FTF conversation because it is less intrusive and allows multitasking. Furthermore, IM has a huge base of users. Market reports indicate that over 140M people worldwide used IM at the end of 2000 (Radicati Group, 2001). In addition, IM is used by multiple populations for different purposes and It has been widely adopted by teenagers for socializing, and by adults for both social and work purposes.

**Literature Review**

Effective communication is a critical component of successful collaboration. It enables collaborators to foster ideas, to build common ground, and to develop complex interpersonal relationships (Tsai & Ghoshal, 1998). As new communication technologies emerge, their use is becoming increasingly common in the workplace. The office is no longer just telephone, email and FAX. CSCW researchers have studied successful use and adoption of instant messaging/chat (Kim, Gwang., Park, & Rice, 2007) virtual worlds, social networking sites (Skeels & Grudin, 2009), Twitter (Zhao & Rosson, 2004), wikis and blogs (Danis & Singer, 2008) in the workplace, and have found them to be beneficial.

Developed in the 1990s for personal chat and entertainment, instant messaging (IM) is rapidly becoming a de facto standard for instantaneous communications within the workplace. Recent research indicates that more than 85 percent of all businesses and organisation now make use of IM. Additionally, one in three IM users now utilize IM as much or more than e-mail, and many predict that IM usage will outstrip e-mail usage within the next few years (Quest Software Inc, 2008). Instant messaging, or IM as it's commonly referred to, is a form of text-based, real-time communication, carried out between two or more people over a digital network. Most people use their personal computers for instant messaging conversations over the Internet, but these chats are also becoming more and frequent on mobile devices over cellular networks (Hedlund, 2011).

Further Rouse (2003) described instant messaging as an Internet service that allows the user to communicate in real time with other users who have the same instant messaging application. Instant messaging includes something called presence technology, which means that when the user launches the application, they can see who on their contact list is online. Icons on the contact list also indicate who is online but not available for instant messaging, and whether or not the contact is using a mobile device (Rouse, 2003).

Instant messages are basically a chat room for two and conversations flow rather like a telephone conversation; even during peak Internet usage periods, the delay is rarely more than a second or two. In addition to allowing the user to send either text or voice messages, many instant messaging services permit the sharing of Web links, images, sounds, streaming content and files. Most instant messaging applications also permit group chats (Rouse, 2003). Instant messaging falls into a category of IT called groupware; meaning programs that help people work together collectively while located remotely from each other.

According to Jones, (2013) instant messaging now allows co-workers to get their ideas across in real time. This ability to communicate quickly, privately, and on the fly makes instant messaging one of the best new tools available in the workplace. Instant messaging is great when communicating with clients and colleagues. Succeeding in the workplace now often involves being able to use instant messaging to its full advantage. Use of the technology, which allows for synchronous, virtual communication, has been steadily rising over the past five years. (Madden 2003) Instant Messaging, also known as online chat, represents the most impressive online revolution since the advent of email.

In order to fully utilize instant messaging in the workplace it is important to be brief, appropriate and understandable. Brief messages are easy to understand and can be read quickly. By creating brief messages the conversation can flow easily between clients and colleagues. This allows business to be settled quickly and efficiently. It's also important to be appropriate while instant messaging at the workplace, especially when messaging other workers. Keep messages limited to business or to small talk. Avoid any messages that may be considered unprofessional (Jones, 2013).

Instant messaging applications are generally categorized as either being public or enterprise. AOL's instant messenger (AIM), Yahoo Messenger and Microsoft .NET Messenger are examples of public IM services. Anyone on the Internet can sign up, download the software and begin messaging. Sun ONE Instant Messaging, IBM Lotus Instant Messaging & Web Conferencing and Microsoft Office Live Communications Server 2003 are examples of enterprise IM services. Access to the IM server is restricted and security precautions, such as encryption, are put in place to protect the enterprise network.

Instant messaging differs from ordinary e-mail in the immediacy of the message exchange and also makes a continued exchange simpler than sending e-mail back and forth. Most exchanges are text only, though popular services, such as AOL, MSN Messenger, Yahoo! Messenger and Apple's iChat now allow voice messaging, file sharing and even video chat when both users have cameras.

**Instant messaging Concept**

Most instant messaging systems work the same way. When one launches the application, the messaging client attempts to connect to the messaging server. The messaging server verifies the username and password and logs the client on. Once it's logged on, the client sends the server its IP address, the port number that's been assigned to the IM service and the names of everyone on the user's contact list. The server creates a temporary session file that contains the connection information and checks to see who on the contact list is also logged on. Because the client has the IP address and port number for the computer of the person that the message was sent to the message is sent directly to the client on that person's computer. In other words, the server is not involved at this point. All communication is directly between the two clients (Rouse, 2003; Jeff & Cooper, 2001).

When the server finds contacts that are logged on, it sends a message back to the client with their connection information and sends the connection information to the contacts. As soon as all the connection information has been sent and acknowledged, instant messaging can begin. The connection process generally takes about ten seconds. The other person gets your instant message and responds. The window that each of one sees on the respective computers expands to include a scrolling dialog of the conversation. Each person's instant messages appear in this window on both computers (Rouse, 2003).

When the conversation is complete, the message window is closed. Eventually, one goes offline and exit. When this happens, the client sends a message to the server to terminate the session. The server sends a message to the client of each person on the contact list who is currently online to indicate that one has logged off. Finally, the server deletes the temporary file that contained the connection information for your client. In the clients of your contacts that are online, your name moves to the offline status section (Jeff & Cooper, 2001).

Instant messaging has been widely used with the power of internet, people can use an IM talk to family, friends, co-workers, even make new friends, join a interesting discussion or chat room through internet, in this way, people can talk to anyone in the world. As a rule for almost everything, while there are advantages, there are also downsides to instant messenger or IMs. In this article, we will take a look at the advantages and the disadvantages of this instant technology.

**Advantages**

There are many advantages to using instant messenger. It connects people regardless of where they are actually located. In the company, colleagues can send and reply instant message in real time without face to face, meanwhile the work report can be shared during the instant chat session; the IM can make a virtual conference without get all the related people together in a physical meeting room. People can speak to multiple people in the virtual conference, share ideas and get conclusions. People on a business trip can contact the co-works inside the company through IM as well. What's more, the staff can talk to customers or vendors online as well, in other word, now people can do business through the instant messenger direct rather than use the traditional method like make phone calls and sending mails (Mahmood, 2013).

IM has proven return on investment benefits in certain situations, such as conferencing. Most IM clients make it easy for several people to participate in the same discussion, at amuch lower cost and with less hassle than setting up a phone conference. Group members can conference in to such a conversation from around the world saving on long distance charges and travel expenses. IM meetings also tend to be more efficient and less prone to straying off topic, because of the relative effort of typing versus talking (Osterman Research, 2006).

A study by the Radicati Group (2004) looked at the time it took employees to complete two typical daily tasks both with and without IM and found that companies could save an average of 40 minutes a day per user with IM. They estimated that an organization with 5,000 people could see a $37.5 million a year savings in productivity. While those estimates didn't factor in the additional costs of managing security and compliance issues, Radicati group (2004) is confident that organizations would nevertheless reap significant Return on Investment improvements from IM deployment.

In their research, Quest Software Inc (2008) noted that Workplace use of IM provides a host of benefits within the organization. Its presence features and immediacy can eliminate much of the internal churn and waste of e-mail, voice mail or office visits. It provides contact with remote employees, customers and vendors at a more intimate level than other forms of electronic communication. Overall IM boosts business performance by making operations faster, more agile, and more efficient with very little additional cost.

IM is less intrusive than a phone call, but more immediate than email. And it has the added advantage of being able to detect presence. Users can set status messages telling others whether they are available or not, which adds to IM's value as a skilful means of communication. IM offers a way to quickly resolve questions and issues as they arise, and managers open to using IM find it an essential medium for receiving feedback and information from their staff (Perey, 2004).

Perey, (2004) has noted that IM has proven its overwhelming value when it comes to gathering input from many different people in dispersed locations. Processes that were once agonizingly slow and inclined toward misunderstanding and errors can now be accomplished in record time. When questions arise, the telephone is no longer an obstacle. In fact, many people use IM and the phone simultaneously. Company employees can chat with each other privately while on a group call with an external partner, for example. Instant messaging is the multi-tasking tool of choice most people chatting over IM does other things at the same time. According to Hedlund( 2011) supported this argument Multi-tasking in this sense to mean that people can engage in instant messaging conversations while in a phone call or while going through their email inbox.

Instant messaging by nature encourages people to be brief and to the point. Increase in productivity is gained by eliminating time wasted switching between various communication methods such as sending and waiting for a response to an email, calling someone and leaving a voice mail when there is no response, walking down the office to see if someone is available only to find that they are in a meeting, and so on (Hedlund, 2011).

**Disadvantages**

There are of course disadvantages of instant messaging. While the real time response is great, IMs do take away the face to face, personal experience that people have when they are speaking to someone in person. You can't really get a very good emotional bead on someone through IM nor are you sure of who you are talking to through IM, which can be dangerous if you aren't careful (Woodard, 2011). Hedlund (2011) noted the downside of IM that is sometimes raised is the loss of relationship building when face to face meetings and phone calls are being replaced with short, text-based conversations.

Hedland(2011) further noted that Some of the reasons why executives are against instant messaging is the increased risk of company confidential information leaking, increase in workplace gossip, loss of work time due to personal instant messaging conversations, potential legal claims, regulatory fines, and security breaches. Uncertainty that the person you are talking to is the person you are talking to, especially when you are not very familiar with the things and risks of the internet. This can be dangerous. Also your computer may be attacked of viruses due to you may accidentally receive some files from the unknown people or click a disguised URL (Woodard, 2011).

One of the most serious threats to businesses that allow unmonitored IM use involves the potential loss of confidential data either proprietary business information or sensitive customer or employee data. Public IM clients don't usually include an encryption option, so any information shared in an IM conversation has the potential of being intercepted. Employees may be sitting across the room from each other, but the IM messages that pass between them are leaving and re-entering the network passing through the corporate firewall, out into the "cloud" that is the Internet, and back (Mahmood, 2013).

In company environment, there will be potential for misuse. People in workplaces may use the IMs during work time to chat with friends and waste time or even bring the possible virus from outside (Mahmood, 2013). There are also some security risks like the content of the instant message may be intercepted or disclose information when conducting several different IM conversations at once, and accidentally send a message to the wrong person. As a result the sensitive data like customer list, sales report may be revealed on the internet.

Lack of clear ICT policies and regulations that governs the use of instant messaging is also a major hindrance to the effective use of the IM in the work place. Many organizations install and roll out instant messaging capabilities to their employees without having any formal policies or guidelines in place, and without providing training for their users. Since many people are familiar with instant messaging for personal, social communication, it is a major risk that they apply the same form of communication in the workplace, which may or may not be appropriate (Hedland, 2011).

**Conclusion**

In order for institution to maximize increased business productivity using instant messaging it's imperative that organizations define and publish ICT policies, guidelines and regulations. Employees need to be trained and be made aware of the ICT Policies that the company infrastructure should be used for work-related communications. Limited form of personal instant messaging might increase business productivity since it is much quicker and less disruptive than a phone call. In addition to following policies and guidelines, it is also very important that employees use good instant messaging etiquette as defined by the policies.

IM use in work place is less interfering than a phone call and more immediate than an email, it has the added advantage of being able to detect presence. Users can set status messages telling others whether they are available or not, which adds to IM's value as a skilful means of communication. IM offers a way to quickly resolve questions and issues as they arise, and managers should adopt IM as an essential medium for receiving feedback and information from their staff.

Organizations that deploy IM would reap significant Return on Investment, IM has proven return on investment benefits in certain situations, at a much lower cost and with less hassle thus saving on long distance charges and travel expenses. Institutions should adopt IM meetings which are be more efficient and less prone to straying off topic, because of the relative effort of typing versus talking.